\documentclass[12pt]{article}
\usepackage{theorem}
\usepackage{amssymb}
\usepackage{graphicx}
\usepackage[mathscr]{eucal}
{\theorembodyfont{\upshape} }
\newcommand{\boma}[1]{{\mbox{\boldmath $#1$} }}

\begin{document}
\newcommand{\binom}[2]{\left( \barray{c} #1 \\ #2 \farray \right)}
\newcommand{\uper}[1]{\stackrel{\barray{c} {~} \\ \mbox{\footnotesize{#1}}\farray}{\longrightarrow} }
\newcommand{\nop}[1]{ \|#1\|_{\piu} }
\newcommand{\no}[1]{ \|#1\| }
\newcommand{\nom}[1]{ \|#1\|_{\meno} }
\newcommand{\UU}[1]{e^{#1 \AA}}
\newcommand{\UD}[1]{e^{#1 \Delta}}
\newcommand{\bb}[1]{\mathbb{{#1}}}
\newcommand{\HO}[1]{\bb{H}^{{#1}}}
\newcommand{\Hz}[1]{\bb{H}^{{#1}}_{\zz}}
\newcommand{\Hs}[1]{\bb{H}^{{#1}}_{\ss}}
\newcommand{\Hg}[1]{\bb{H}^{{#1}}_{\gam}}
\newcommand{\HM}[1]{\bb{H}^{{#1}}_{\so}}
\newcommand{\vers}[1]{\widehat{#1}}
\def\tvainf{\vspace{-0.4cm} \barray{ccc} \vspace{-0,1cm}{~}
\\ \vspace{-0.2cm} \longrightarrow \\ \vspace{-0.2cm} \scriptstyle{T \vain + \infty} \farray}
\def\gm{\eta}
\def\Sz{S_z}
\def\Bi{\mathscr{B}}
\def\xu{x^1}
\def\xd{x^2}
\def\xt{x^3}
\def\xtp{{x^3}'}
\def\ki{k_i}
\def\ku{k_1}
\def\kd{k_2}
\def\kt{k_3}
\def\hu{h_1}
\def\hd{h_2}
\def\htt{h_3}
\def\qu{q_1}
\def\qd{q_2}
\def\qt{q_3}
\def\ha{\widehat{a}}
\def\ak{\ha_k}
\def\ah{\ha_h}
\def\had{\ha^{\dagger}}
\def\akd{\had_k}
\def\ahd{\had_h}
\def\effe{F}
\def\Fk{\effe_k}
\def\Fh{\effe_h}
\def\Fkc{\overline{\Fk}}
\def\Fhc{\overline{\Fh}}
\def\fk{f_k}
\def\fh{f_h}
\def\fkc{\overline{\fk}}
\def\fhc{\overline{\fh}}
\def\bx{{\bf x}}
\def\Nab{\square}
\def\Fi{\widehat{\phi}}
\def\s{u}
\def\Fis{\Fi^{\s}}
\def\Ti{\widehat{T}}
\def\Tis{\Ti^{\s}}
\def\Aa{\widehat{A}}
\def\Bb{\widehat{B}}
\def\tu{\xi}
\def\ep{\xi}
\def\iep{(-1,+\infty)}
\def\Do{\mathscr{E}}
\def\CA{\mathcal{C}}
\def\CB{\mathcal{D}}
\def\ca{c}
\def\cb{d}
\def\op{\,\mbox{\scriptsize{or}}\,}
\def\er{\epsilon}
\def\erd{\er_0}
\def\vk{\vers{k}}
\def\vh{\vers{h}}
\def\Qn{\mathfrak{K}_n}
\def\Ed{\hat{E}}
\def\um{u_{-}}
\def\up{u_{+}}
\def\el{t}
\def\em{z}
\def\uu{\lambda}
\def\dK{\delta {\mathscr K}}
\def\dG{\delta {\mathscr G}}
\def\DK{\Delta {\mathscr K}}
\def\DG{\Delta {\mathscr G}}
\def\Km{{\mathscr K}}
\def\Ll{\mathscr{L}}
\def\Hh{\mathscr{H}}
\def\Mm{{\mathscr M}}
\def\Nn{{\mathscr N}}
\def\Rr{{\mathscr R}}
\def\Gg{{\mathscr G}}
\def\Zz{Z}
\def\Ss{{\mathscr S}}
\def\Fe{{\mathscr F}}
\def\Ei{{\mathscr E}}
\def\Ww{{\mathscr W}}
\def\we{\wedge}
\def\We{\bigwedge}
\def\dbar{\hat{d}}
\def\Cc{\mathscr{C}}
\def\SZ{\mathcal{S}}
\def\TZ{\mathfrak{S}}
\def\CQ{S}
\def\GQ{T}
\def\C{{C_d}}
\def\Ac{\overline{A}}
\def\Bc{\overline{B}}
\def\Xd{\Zd_{0 k} \cap \Bd_{0 k}(\ro)}
\def\Yd{\Zd_{0 k} \setminus \Bd_{0 k}(\ro) }
\def\comple{\scriptscriptstyle{\complessi}}
\def\nume{0.407}
\def\numerob{0.00724}
\def\deln{7/10}
\def\delnn{\dd{7 \over 10}}
\def\e{c}
\def\p{p}
\def\z{z}
\def\symd{{\mathfrak S}_d}
\def\Del{\delta}
\def\Di{\Delta}
\def\mmu{\hat{\mu}}
\def\rot{\mbox{rot}\,}
\def\curl{\mbox{curl}\,}
\def\XS{\boma{x}}
\def\TS{\boma{t}}
\def\Lam{\boma{\eta}}
\def\DS{\boma{\rho}}
\def\KS{\boma{k}}
\def\LS{\boma{\lambda}}
\def\PR{\boma{p}}
\def\VS{\boma{v}}
\def\ski{\! \! \! \! \! \! \! \! \! \! \! \! \! \!}
\def\h{L}
\def\EM{M}
\def\EMP{M'}
\def\R{R}
\def\E{E}
\def\FFf{\mathscr{F}}
\def\A{F}
\def\Xim{\Xi_{\meno}}
\def\Ximn{\Xi_{n-1}}
\def\lan{\lambda}
\def\om{\omega}
\def\Om{\Omega}
\def\Oma{\Om_a}
\def\Omp{\Om_{\infty}}
\def\Sim{\Sigm}
\def\Sip{\Delta \Sigm}
\def\Sigm{{\mathscr{S}}}
\def\Ki{{\mathscr{K}}}
\def\Hi{{\mathscr{H}}}
\def\zz{{\scriptscriptstyle{0}}}
\def\ss{{\scriptscriptstyle{\Sigma}}}
\def\gam{{\scriptscriptstyle{\Gamma}}}
\def\so{\ss \zz}
\def\Dz{\bb{\DD}'_{\zz}}
\def\Ds{\bb{\DD}'_{\ss}}
\def\Dsz{\bb{\DD}'_{\so}}
\def\Dg{\bb{\DD}'_{\gam}}
\def\Ls{\bb{L}^2_{\ss}}
\def\Lg{\bb{L}^2_{\gam}}
\def\bF{{\bb{V}}}
\def\Fz{\bF_{\zz}}
\def\Fs{\bF_\ss}
\def\Fg{\bF_\gam}
\def\Pre{P}
\def\UUU{{\mathcal U}}
\def\fiapp{\phi}
\def\PU{P1}
\def\PD{P2}
\def\PT{P3}
\def\PQ{P4}
\def\PC{P5}
\def\PS{P6}
\def\Q{P6}
\def\X{Q2}
\def\Xp{Q3}
\def\Vi{V}
\def\bVi{\bb{V}}
\def\K{V}
\def\Ks{\bb{\K}_\ss}
\def\Kz{\bb{\K}_0}
\def\KM{\bb{\K}_{\, \so}}
\def\HGG{\bb{H}^\G}
\def\HG{\bb{H}^\G_{\so}}
\def\EG{{\mathfrak{P}}^{\G}}
\def\G{G}
\def\de{\delta}
\def\esp{\sigma}
\def\dd{\displaystyle}
\def\LP{\mathfrak{L}}
\def\dive{\mbox{div}}
\def\la{\langle}
\def\ra{\rangle}
\def\um{u_{\meno}}
\def\uv{\mu_{\meno}}
\def\Fp{ {\textbf F_{\piu}} }
\def\Ff{ {\textbf F} }
\def\Fm{ {\textbf F_{\meno}} }
\def\piu{\scriptscriptstyle{+}}
\def\meno{\scriptscriptstyle{-}}
\def\omeno{\scriptscriptstyle{\ominus}}
\def\Tt{ {\mathscr T} }
\def\Xx{ {\textbf X} }
\def\Yy{ {\textbf Y} }
\def\VP{{\mbox{\tt VP}}}
\def\CP{{\mbox{\tt CP}}}
\def\cp{$\CP(f_0, t_0)\,$}
\def\cop{$\CP(f_0)\,$}
\def\copn{$\CP_n(f_0)\,$}
\def\vp{$\VP(f_0, t_0)\,$}
\def\vop{$\VP(f_0)\,$}
\def\vopn{$\VP_n(f_0)\,$}
\def\vopdue{$\VP_2(f_0)\,$}
\def\leqs{\leqslant}
\def\geqs{\geqslant}
\def\mat{{\frak g}}
\def\tG{t_{\scriptscriptstyle{G}}}
\def\tN{t_{\scriptscriptstyle{N}}}
\def\TK{t_{\scriptscriptstyle{K}}}
\def\CK{C_{\scriptscriptstyle{K}}}
\def\CN{C_{\scriptscriptstyle{N}}}
\def\CG{C_{\scriptscriptstyle{G}}}
\def\CCG{{\mathscr{C}}_{\scriptscriptstyle{G}}}
\def\tf{{\tt f}}
\def\ti{{\tt t}}
\def\ta{{\tt a}}
\def\tc{{\tt c}}
\def\tF{{\tt R}}
\def\P{{\mathscr P}}
\def\V{{\mathscr V}}
\def\TI{\tilde{I}}
\def\TJ{\tilde{J}}
\def\Lin{\mbox{Lin}}
\def\Hinfc{ H^{\infty}(\reali^d, \complessi) }
\def\Hnc{ H^{n}(\reali^d, \complessi) }
\def\Hmc{ H^{m}(\reali^d, \complessi) }
\def\Hac{ H^{a}(\reali^d, \complessi) }
\def\Dc{\DD(\reali^d, \complessi)}
\def\Dpc{\DD'(\reali^d, \complessi)}
\def\Sc{\SS(\reali^d, \complessi)}
\def\Spc{\SS'(\reali^d, \complessi)}
\def\Ldc{L^{2}(\reali^d, \complessi)}
\def\Lpc{L^{p}(\reali^d, \complessi)}
\def\Lqc{L^{q}(\reali^d, \complessi)}
\def\Lrc{L^{r}(\reali^d, \complessi)}
\def\Hinfr{ H^{\infty}(\reali^d, \reali) }
\def\Hnr{ H^{n}(\reali^d, \reali) }
\def\Hmr{ H^{m}(\reali^d, \reali) }
\def\Har{ H^{a}(\reali^d, \reali) }
\def\Dr{\DD(\reali^d, \reali)}
\def\Dpr{\DD'(\reali^d, \reali)}
\def\Sr{\SS(\reali^d, \reali)}
\def\Spr{\SS'(\reali^d, \reali)}
\def\Ldr{L^{2}(\reali^d, \reali)}
\def\Hinfk{ H^{\infty}(\reali^d, \KKK) }
\def\Hnk{ H^{n}(\reali^d, \KKK) }
\def\Hmk{ H^{m}(\reali^d, \KKK) }
\def\Hak{ H^{a}(\reali^d, \KKK) }
\def\Dk{\DD(\reali^d, \KKK)}
\def\Dpk{\DD'(\reali^d, \KKK)}
\def\Sk{\SS(\reali^d, \KKK)}
\def\Spk{\SS'(\reali^d, \KKK)}
\def\Ldk{L^{2}(\reali^d, \KKK)}
\def\Knb{K^{best}_n}
\def\sc{\cdot}
\def\k{\mbox{{\tt k}}}
\def\g{ {\textbf g} }
\def\QQQ{ {\textbf Q} }
\def\AAA{ {\textbf A} }
\def\gr{\mbox{gr}}
\def\sgr{\mbox{sgr}}
\def\loc{\mbox{loc}}
\def\PZ{{\Lambda}}
\def\PZAL{\mbox{P}^{0}_\alpha}
\def\epsilona{\epsilon^{\scriptscriptstyle{<}}}
\def\epsilonb{\epsilon^{\scriptscriptstyle{>}}}
\def\lgraffa{ \mbox{\Large $\{$ } \hskip -0.2cm}
\def\rgraffa{ \mbox{\Large $\}$ } }
\def\restriction{\upharpoonright}
\def\M{{\scriptscriptstyle{M}}}
\def\m{m}
\def\Fre{Fr\'echet~}
\def\I{{\mathcal N}}
\def\ap{{\scriptscriptstyle{ap}}}
\def\fiap{\varphi_{\ap}}
\def\dfiap{{\dot \varphi}_{\ap}}
\def\DDD{ {\mathfrak D} }
\def\BBB{ {\textbf B} }
\def\EEE{ {\textbf E} }
\def\GGG{ {\textbf G} }
\def\TTT{ {\textbf T} }
\def\KKK{ {\textbf K} }
\def\HHH{ {\textbf K} }
\def\FFi{ {\bf \Phi} }
\def\GGam{ {\bf \Gamma} }
\def\sc{ {\scriptstyle{\bullet} }}
\def\a{a}
\def\c{\kappa}
\def\parn{\par\noindent}
\def\teta{M}
\def\elle{L}
\def\ro{\rho}
\def\al{\alpha}
\def\alc{\overline{\al}}
\def\dal{\mathfrak{a}}
\def\si{\sigma}
\def\be{\beta}
\def\dbe{\mathfrak{b}}
\def\bec{\overline{\be}}
\def\dbec{\overline{\dbe}}
\def\ga{\gamma}
\def\tet{\vartheta}
\def\teta{\theta}
\def\ch{\chi}
\def\et{\eta}
\def\complessi{{\bf C}}
\def\len{{\bf L}}
\def\reali{{\bf R}}
\def\interi{{\bf Z}}
\def\Z{{\bf Z}}
\def\naturali{{\bf N}}
\def\To{ {\bf T} }
\def\Td{ {\To}^d }
\def\Tt{ {\To}^3 }
\def\Bd{B^d}
\def\Zd{ \interi^d }
\def\Zt{ \interi^3 }
\def\Zet{{\mathscr{Z}}}
\def\Ze{\Zet^d}
\def\T1{{\textbf To}^{1}}
\def\Sfe{ {\bf S} }
\def\Sd{\Sfe^{d-1}}
\def\St{\Sfe^{2}}
\def\es{s}
\def\FF{\mathcal F}
\def\FFu{ {\textbf F_{1}} }
\def\FFd{ {\textbf F_{2}} }
\def\GG{{\mathcal G} }
\def\EE{{\mathcal E}}
\def\KK{{\mathcal K}}
\def\PP{{\mathcal P}}
\def\PPP{{\mathscr P}}
\def\PN{{\mathcal P}}
\def\PPN{{\mathscr P}}
\def\QQ{{\mathcal Q}}
\def\J{J}
\def\Np{{\hat{N}}}
\def\Lp{{\hat{L}}}
\def\Jp{{\hat{J}}}
\def\Vp{{\hat{V}}}
\def\Ep{{\hat{E}}}
\def\Gp{{\hat{G}}}
\def\Kp{{\hat{K}}}
\def\Ip{{\hat{I}}}
\def\Tp{{\hat{T}}}
\def\Mp{{\hat{M}}}
\def\La{\Lambda}
\def\Ga{\Gamma}
\def\Si{\Sigma}
\def\Upsi{\Upsilon}
\def\Gam{\Gamma}
\def\Gag{{\check{\Gamma}}}
\def\Lap{{\hat{\Lambda}}}
\def\Upsig{{\check{\Upsilon}}}
\def\Kg{{\check{K}}}
\def\ellp{{\hat{\ell}}}
\def\j{j}
\def\jp{{\hat{j}}}
\def\BB{{\mathcal B}}
\def\LL{{\mathcal L}}
\def\MM{{\mathcal U}}
\def\SS{{\mathcal S}}
\def\DD{D}
\def\Dd{{\mathcal D}}
\def\VV{{\mathcal V}}
\def\WW{{\mathcal W}}
\def\OO{{\mathcal O}}
\def\RR{{\mathcal R}}
\def\TT{{\mathcal T}}
\def\AA{{\mathcal A}}
\def\CC{{\mathcal C}}
\def\JJ{{\mathcal J}}
\def\NN{{\mathcal N}}
\def\HH{{\mathcal H}}
\def\XX{{\mathcal X}}
\def\XXX{{\mathscr X}}
\def\YY{{\mathcal Y}}
\def\ZZ{{\mathcal Z}}
\def\cir{{\scriptscriptstyle \circ}}
\def\circa{\thickapprox}
\def\vain{\rightarrow}
\def\parn{\par \noindent}
\def\salto{\vskip 0.2truecm \noindent}
\def\spazio{\vskip 0.5truecm \noindent}
\def\vs1{\vskip 1cm \noindent}
\def\fine{\hfill $\square$ \vskip 0.2cm \noindent}
\def\ffine{\hfill $\lozenge$ \vskip 0.2cm \noindent}
\newcommand{\rref}[1]{(\ref{#1})}
\def\beq{\begin{equation}}
\def\feq{\end{equation}}
\def\beqq{\begin{eqnarray}}
\def\feqq{\end{eqnarray}}
\def\barray{\begin{array}}
\def\farray{\end{array}}
\makeatletter \@addtoreset{equation}{section}
\renewcommand{\theequation}{\thesection.\arabic{equation}}
\makeatother
\begin{titlepage}
{~}
\vspace{-2cm}
\begin{center}
{\huge Local zeta regularization and the Casimir effect}
\end{center}
\vspace{0.5truecm}
\begin{center}
{\large
Davide Fermi$\,{}^a$, Livio Pizzocchero$\,{}^b$({\footnote{Corresponding author}})} \\
\vspace{0.5truecm}
${}^a$ Universit\`a di Milano \\
e--mail: davide.fermi@studenti.unimi.it \\
\vspace{0.2truecm}
${}^b$ Dipartimento di Matematica, Universit\`a di Milano\\
Via C. Saldini 50, I-20133 Milano, Italy\\
and Istituto Nazionale di Fisica Nucleare, Sezione di Milano, Italy \\
e--mail: livio.pizzocchero@unimi.it
\end{center}
\begin{abstract}
In this paper, whose aims are mainly pedagogical, we illustrate
how to use the local zeta regularization
to compute the stress-energy tensor of the Casimir
effect. Our attention is devoted to the case of
a neutral, massless scalar field in flat space-time, on
a space domain with suitable (e.g., Dirichlet)
boundary conditions. After
a simple outline of the local zeta method, we
exemplify it in the typical case of
a field between two parallel plates, or outside them.
The results are shown to agree with the ones
obtained by more popular methods, such as
point splitting regularization. In comparison
with these alternative methods, local zeta regularization
has the advantage to give directly finite results
via analitic continuation, with no need to remove
or subtract divergent quantities.
\end{abstract}
\vspace{0.2cm} \noindent
\hfill \parn
\par \vspace{0.3truecm} \noindent \textbf{PACS}: 03.70.+k, 11.10.Gh~.
\end{titlepage}
\section{Introduction}
\label{intro}
Zeta regularization is a method to give meaning to the divergent series
appearing frequently in quantum field theory, reinterpreting them as analytic
continuations. For example, the divergent series
\beq \sum_{\ell=1}^{\infty} \ell^3 \label{divser} \feq
is interpreted in this approach as the analytic continuation at $s=-3$ of
the regularized series $\zeta(s) := \sum_{\ell=1}^{+\infty} 1/\ell^s$, that
converges for $\Re s > 1$ and defines the familiar Riemann zeta function;
in this sense, the sum \rref{divser} ''equals'' $\zeta(-3) = 1/120$. Tricks
of this kind have been used for a long time in quantum field theory: for example,
the above analysis of the series \rref{divser} appears in one of the most
popular derivations of the total Casimir energy for a scalar or electromagnetic field
between two parallel plates (see \cite{Blau, Zeta, Mil}, or
the issue ``Casimir effect'' in Wikipedia). \parn
The computation of local quantities, such as (the vacuum expectation value of)
the stress-energy tensor, can be done as well via a generalization
of the above method; this procedure, called the \textsl{local zeta regularization},
is a bit less popular than its analogue for global quantities,
such as the total energy. \parn
The method of local zeta regularization arose from some ideas of
Hawking \cite{Hawk} and Wald \cite{Wal}; these were systematically developed
and applied to the stress-energy tensor by Moretti
in a long series of papers, among which we quote \cite{Mor1,Mor2}.
\parn
These authors typically work on curved space-times, in a Euclidean framework (i.e., with a
space-time metric of signature $(+,+,+,+)$) (indeed, \cite{Hawk} \cite{Mor1}
also mention a flate case, namely, a four dimensional Euclidean torus);
in this setting, the local zeta regularization is applied
to divergent sums arising from path integrals. \parn
Since the local zeta method is not so popular, in our opinion it is not
useless, at least for pedagogical reasons, to illustrate it
in a much simpler framework; this is the aim of the present work. \parn
In this paper we consider a (neutral, massless)
scalar field $\Fi$ in Minkowski space-time (with the usual
metric of signature $(-,+,+,+)$), as viewed
in a given inertial frame; the field is canonically quantized
on a three-dimensional space domain $\Omega$, with
suitable boundary conditions on the frontier $\partial \Omega$.
We are interested in the stress-energy tensor $\Ti_{\mu \nu}$ or,
more precisely, in the vacuum expectation value (VEV) of each stress-energy component:
this is the so-called \textsl{local problem}
in the theory of the Casimir effect. \parn
To deal with the divergences related to this problem, we introduce
a ''zeta regularized field'' $\Fi^u$, depending on
a complex parameter $u$ and coinciding with $\Fi$
for $u=0$; this is used to define a ''zeta regularized stress-energy tensor''
$\Tis_{\mu \nu}$, with finite VEV. The final step in this construction is the analytic
continuation at $u=0$ of such VEV, which give the Casimir
stress-energy for the case under consideration. A very pleasant
feaure of this approach, typical of the zeta regularization method,
is that one gets directly a finite expression
for the Casimir stress-energy, with no need to remove or
subtract divergent terms. This is a major difference with respect to
other renormalization schemes, employed more frequently
for the Casimir effect; among these alternative approaches,
let us mention the point splitting method, occasionally considered in this paper
for a comparison. \parn
In the present work, the local zeta approach is mainly illustrated
in the case of a field between two parallel plates,
with Dirichlet boundary conditions (so, in this example
$\Omega = (-\infty, + \infty)^2 \times (0,a)$, with $a$
the distance between the plates); by simple
variations of this setting, we then pass to the case
of a field in the region outside one or two plates.
The results obtained
in these cases by local zeta regularization are compared with the ones
derived by Milton \cite{Mil} and by Esposito \textsl{et al}
\cite{Esp} by point splitting, and they are found to coincide
(incidentally, we take the occasion to show that
the Casimir stress-energy tensors found in \cite{Mil}
and \cite{Esp}, even though seemingly different, are
in fact equal). \parn
To conclude this Introduction, let us briefly
describe the organization of the paper.
In Section \ref{back} we sketch the basic
framework for the Casimir effect, in the situation
outlined before: a canonically quantized scalar field on a space domain
$\Omega$ in Minkowski space-time, with given
boundary conditions, its stress-energy tensor and the related
VEV. In Section \ref{locaze} we introduce the local
zeta regularization scheme. In Section \ref{capar},
we apply this scheme to the case of
a Dirichlet field bewteen parallel plates, giving all details
about the necessary analytic continuations; comparison
is made with the results of \cite{Mil} \cite{Esp}. In Section
\ref{caout}, by simple geometric variations
on the same theme, we derive the Casimir stress-energy tensor
outside one plate, or two parallel plates.
In Section \ref{press}, the outcomes of
Sections \ref{capar}, \ref{caout} are combined to derive
the Casimir pressure on two parallel plates. \parn
For completeness, in Appendices \ref{copoly} and \ref{copoly3} we give some
mathematical background on the analytic continuation
of the polylogarithm, the function mainly involved
in local zeta regularization for the case under investigation.
\salto
\section{Background for the scalar Casimir effect.}
\label{back}
Throughout this note we work in Minkowski space-time, which is
identified with $\reali^4$ using a set of inertial coordinates
\beq x = (x^\mu)_{\mu=0,1,2,3} \equiv (t, \xu, \xd, \xt) \equiv (t, \bx)~. \feq
We work in units where $c=1, \hbar = 1$; the Minkowski metric is
$(\gm_{\mu \nu}) = \mbox{diag}(-1,1,1,1)$ (and is used
to raise and lower indices). \parn
Let us fix a space domain $\Om \subset \reali^3$ where we consider
a neutral, massless scalar field $\Fi$; so, we have
\beq \Fi : \reali \times \Om \mapsto \LL_{s a}(\HH)~; \qquad 0 =
\Nab \Fi = (- \partial_{t t} + \Delta) \Fi~. \label{daquan} \feq
Here we are considering the space $\LL(\HH)$ of linear operators on the Fock space $\HH$,
and the subset $\LL_{s a}(\HH)$ of the selfadjoint operators;
$\Nab := \partial^\mu \partial_{\mu}$ is the d'Alembertian and $\Delta := \sum_{i=1}^3 \partial_{ii}$
is the 3-Laplacian. We assume appropriate boundary conditions
(e.g., the Dirichlet conditions $\Fi(t,\bx) = 0$ for $\bx \in \partial \Omega$). \parn
To expand the field in normal modes, we consider a complete orthonormal set
$(\Fk)_{k \in K}$ of eigenfunctions for the Laplacian in $L^2(\Omega, \complessi)$, with the given boundary conditions;
$K$ is a space of labels, for the moment unspecified, and we write
the eigenvalues in the form $-\om^2_k$. So,
\beq \Fk : \Omega \vain \complessi;
\quad \Delta \Fk= - \om^2_k \Fk~(\om_k > 0); \feq
$$ \int_{\Om} d^3 \bx \Fkc(\bx) \Fh(\bx) = \delta(k, h) \quad (k,h \in K)~. $$
Any eigenvector label $k$ can include different parameters, both discrete and
continuous. We generically write $\int_{K} d k$ to indicate summation over all labels,
(i.e., literal summation over the discrete parameters and
integration over the continuous parameters, with a suitable measure); $\delta(h,k)=\delta(k,h)$
is the Dirac delta function for the labels space $K$ (this reduces to the Kronecker symbol
in the case of discrete parameters).
The functions
\beq \fk : \reali \times \Om \vain \complessi~, \qquad \fk(x) := \Fk(\bx) e^{- i \om_k t} \feq
fulfill $\Nab f_k = 0$, and allow a unique expansion
\beq \Fi(x) = \int_{K} {d k \over \sqrt{2 \om_k}} \left[ \ak \fk(x) + \akd \fkc(x) \right] \label{espan} \feq
(with $^\dag$ indicating the adjoint operator, and $\overline{\phantom{I}}$ the complex conjugate).
The destruction and creation operators $\ak, \akd \in \LL(\HH)$ fulfill the relations
\beq [\ak, \ah] = 0~, \quad [\ak, \ahd] = \delta(h,k)~, \qquad \ak |0 \ra = 0~, \feq
where $|0 \ra \in \HH$ is the vacuum state (of norm $1$).
\parn
Let us pass to the stress-energy tensor.
This depends
on a parameter $\xi \in \reali$, and its components
$\Ti_{\mu \nu} : \reali \times \Omega \vain \LL_{s a}(\HH)$ are given by
\beq \Ti_{\mu \nu} := \left(1 - 2 \xi \right) \partial_\mu \Fi \circ \partial_\nu \Fi
- \left( {1 \over 2} - 2 \xi \right) \gm_{\mu \nu} \partial^\lan \Fi \partial_{\lan} \Fi -
2 \xi \, \Fi \circ \partial_{\mu \nu} \Fi~; \label{tiquan} \feq
in the above, we use the symmetrized operator product $\Aa \circ \Bb := (1/2) (\Aa \Bb + \Bb \Aa)$. \parn
To be precise, a theory involving mereley a scalar field in flat space-time
has a stress-energy tensor as above, with $\xi=0$; the general form
\rref{tiquan}, with an arbitrary $\xi$, can be interpreted
as the Minkowskian limit of the theory of a massless scalar field coupled with gravity,
with $\xi$ as the coupling constant.
({\footnote{In the theory of a classical, massless scalar field
coupled with gravity, the dynamical variables are
the field $\phi$ and the space-time metric $g_{\mu \nu}$. The action functional is
$S[\phi, g_{\mu \nu}] = {1 \over 2} \int d^4 x \sqrt{-g}\Big(\partial^\mu \phi
\partial_{\mu} \phi - R({1 \over 8 \pi} - \xi \phi^2)\Big)$
in units where the gravitational constant is $1$,
where $g := \det(g_{\mu \nu})$ and $R$ is the scalar curvature of the metric:
for more details see, e.g., \cite{Par} page 43.
One imposes the stationarity of
the action  with respect to variations of $\phi$ and $g_{\mu \nu}$:
in this way one gets, respectively, the scalar field equation and
Einstein's equations with the field stress-energy tensor $T_{\mu \nu}$.
One can analyze the almost Minkowskian case where $\phi$
is small and $g_{\mu \nu} = \gm_{\mu \nu} + h_{\mu \nu}$,
with $h_{\mu \nu}$ a small perturbation of the second order in $\phi$.
In this case the scalar field equation has the form
$\Nab \phi = O_2[\phi]$ with $\Nab$
the Minkowski d'Alembertian $\partial^\mu \partial_\mu$,
and the stress-energy tensor is
$T_{\mu \nu} = \left(1 - 2 \xi \right) \partial_\mu \phi \partial_\nu \phi
- \left( {1 \over 2} - 2 \xi \right) \gm_{\mu \nu} \partial^\lan \phi \partial_{\lan} \phi -
2 \xi \, \phi \partial_{\mu \nu} \phi + O_3[\phi]$; here, $O_2$ and $O_3$ indicate
terms of orders $2$ and $3$. Of course, Einstein's equations
relate $h_{\mu \nu}$ to $\phi$.
Neglecting the higher order terms, and quantizing the field, we obtain Eq.s \rref{daquan} \rref{tiquan}.}}). \parn
Other authors have considered the general form
\rref{tiquan} independently of the previous interpretation in terms
of a gravitational coupling; these authors invoke the principle that one can add
to the stress-energy tensor a symmetric tensor with vanishing divergence,
and regard the terms proportional to $\xi$ in \rref{tiquan}
as additions of this kind \cite{Mil}. \parn
To conclude these comments about $\xi$,
we mention that the choice $\xi=1/6$ gives a conformally invariant theory \cite{Par},
where the tensor \rref{tiquan} has vanishing trace. For the above reasons,
the term \textsl{conformal coupling} is usually employed to describe the
case $\xi=1/6$; one also speaks of a \textsl{minimal coupling} to indicate
the case $\xi=0$. \parn
After this digression, we proceed towards the Casimir effect considering
the vacuum expectation value (VEV) of $\Ti_{\mu \nu}$. We use the expansion \rref{espan} for
the field, with the relations $\la 0 | \ak \ah | 0 \ra = 0$,
$\la 0 | \akd \ahd | 0 \ra = 0$, $\la 0 | \akd \ah | 0 \ra = 0$ and
$\la 0 | \ak \ahd | 0 \ra = \delta(k,h)$; in this way, we readily obtain the formal expression
\beq \la 0 | \Ti_{\mu \nu} | 0 \ra =
\int_{K} {d k \over \om_{k}} \, \left[
\Big({1 \over 4} - {\xi \over 2}\Big)
\Big(\partial_\mu \fk \partial_\nu \fkc + \partial_\nu \fk \partial_\mu \fkc\Big)
\right. \label{formal} \feq
$$ \left. \hspace{2.9cm} - \Big({1 \over 4} - \xi\Big) \gm_{\mu \nu} \partial^\lan \fk \partial_{\lan} \fkc
- {\xi \over 2} \Big(\fk \partial_{\mu \nu} \fkc + \fkc \partial_{\mu \nu} \fk\Big) \right] \, ; $$
however, the above integral is divergent and some renormalization procedure is needed.
A standard approach relies on the so-called point splitting regularization
(see \cite{Bro, Chri, Esp}; \cite{Mil} essentially uses the same method). In this approach, in place
of $\Ti_{\mu \nu}(x)$ one considers
\parn
\vbox{
\beq \Ti_{\mu \nu}(x,x') := \left( 1 - 2 \xi \right) \partial_\mu \Fi(x) \circ \partial_\nu \Fi(x')
- \left( {1 \over 2} - 2 \xi \right) \gm_{\mu \nu} \partial^\lan \Fi(x) \circ \partial_{\lan} \Fi(x')  \label{spli1} \feq
$$ \hspace{-2.2cm} - 2 \xi \, \Fi(x) \circ \partial_{\mu \nu} \Fi(x') \, , \qquad (x, x' \in \reali \times \Om) \, , $$
giving formally $\Ti_{\mu \nu}(x)$ in the limit $x' \vain x$. One then defines the renormalized
VEV of $\Ti_{\mu \nu}(x)$ as
\beq \la 0 | \Ti_{\mu \nu}(x) | 0 \ra_{ren} := FP \Big|_{x' \vain x} \la 0 | \Ti_{\mu \nu}(x,x') | 0 \ra~, \label{spli2}
\feq
}
where we have written $FP$ to indicate the ''finite part'' in the limit $x' \vain x$; this means
that one writes down the VEV of $\Ti_{\mu \nu}(x,x')$ and then removes the terms diverging
for $x' \vain x$. ({\footnote{Of course, the concept
of ''finite part'' contains a basic ambiguity,
that must be removed by a precise prescription.
In the case of an electromagnetic field in Minkowski
space-time, a prescription of this type has been given in \cite{Bro}; this approach
could be adapted to the scalar case. An alternative
strategy is to define the finite part in
\rref{spli2} as the $x'\vain x$ limit
of what remains after substracting
from $\la 0 | \Ti_{\mu \nu}(x,x') | 0 \ra$
the analogous VEV for a field
without boundary conditions. For a
critical analysis about these and other
problematic aspects of point splitting, see
\cite{Mor2}.}}) \parn
Hereafter we will describe the alternative
approach considered in this paper, i.e., the local zeta method.
\salto
\section{Local zeta regularization.}
\label{locaze}
In the sequel we keep all the notations of the previous section.
Let us denote with $\s$ a complex parameter and
consider the powers $(-\Delta)^{-\s/4}$, built from the 3-dimensional
Laplacian $\Delta$. From $\Delta \Fk = - \om_k^2 \Fk$
it follows $(-\Delta)^{-\s/4} \Fk = \om_k^{-\s/2} \Fk$,  whence
\beq (-\Delta)^{-\s/4} \fk = \om_k^{-\s/2} \fk~; \feq
there are similar relations for the conjugate functions, starting from
$\Delta \Fkc = - \om_k^2 \Fkc$. \parn
We now introduce the \textsl{smeared}, or \textsl{zeta-regularized} field operators and stress-energy
tensor
\beq \Fis := (-\Delta)^{-\s/4} \Fi~, \feq
\beq \Tis_{\mu \nu} := \left( 1 - 2 \xi \right) \partial_\mu \Fis \circ \partial_\nu \Fis
- \left( {1 \over 2} - 2 \xi \right) \gm_{\mu \nu} \partial^\lan \Fis \partial_{\lan} \Fis -
2 \xi \, \Fis \circ \partial_{\mu \nu} \Fis~, \feq
which formally give $\Fi$ and $\Ti_{\mu \nu}$ in the limit $\s \vain 0$.
Eq. \rref{espan} implies
\beq \Fis(x) = \int_{K} {d k \over \sqrt{2} \, \om^{\s/2 + 1/2}_k} \left[ \ak \fk(x) + \akd \fkc(x) \right]
\label{espans}~; \feq
now, a computation very similar to the one giving Eq. \rref{formal} produces the result
\parn
\vbox{
\beq \la 0 | \Tis_{\mu \nu} | 0 \ra
= \int_{K} {d k \over \om^{\s+ 1}_{k}} \, \left[
\Big({1 \over 4} - {\xi \over 2}\Big)
\Big(\partial_\mu \fk \partial_\nu \fkc + \partial_\nu \fk \partial_\mu \fkc\Big) \right. \label{formals} \feq
$$ \left. \hspace{2.3cm}  - \Big({1 \over 4} - \xi\Big) \gm_{\mu \nu} \partial^\lan \fk \partial_{\lan} \fkc
- {\xi \over 2} \Big(\fk \partial_{\mu \nu} \fkc + \fkc \partial_{\mu \nu} \fk\Big) \right]. $$
}
The above integral typically converges for $\Re \s$ sufficiently large and is an analytic function of $\s$,
a situation that will be exemplified hereafter.
Eq. \rref{formals}, with $\Re \s$ sufficiently large,
is our regularization of the VEV for $\Ti_{\mu \nu}$; we now define the renormalized VEV as
\beq \la 0 | \Ti_{\mu \nu} | 0 \ra_{ren} := AC \Big|_{\s=0} \la 0 | \Tis_{\mu \nu} | 0 \ra~, \feq
where $AC \Big|_{\s=0}$ indicates that one should consider the analytic continuation
of the function  $\s \mapsto \la 0 | \Tis_{\mu \nu} | 0 \ra$, and evaluate it at $\s=0$. In the
next section, the whole procedure will be exemplified in the classical case
where $\Om$ is the region between two parallel plates, with Dirichlet boundary conditions;
in the subsequent section we will treat the region outside one or two plates.
\section{Casimir effect between two parallel plates}
\label{capar}
\textbf{Setting up the problem; the zeta-regularized stress-energy tensor.}
Let the plates occupy the planes $\xt=0$ and $\xt = a$ ($a>0$); the region between the plates is
\beq \Om := \{ (\xu, \xd, \xt)~|~\xu, \xd \in \reali~, 0 < \xt < a \}~. \feq
We assume the Dirichlet boundary conditions
\beq \Fi(t,\xu, \xt, \xt) = 0 \qquad \mbox{for $\xt=0,a$}~. \feq
Let us produce a complete orthonormal set $(F_k)_{k \in K}$ of Dirichlet eigenfunctions for $\Delta$ on $\Omega$,
and the corresponding eigenvalues $-\om^2_k$. We can take
\beq K := \{ k = (\ku, \kd, \kt)~|~\ku, \kd \in \reali~, \kt \in \{ \pi/a, 2 \pi/a, 3 \pi/a,...\} \}~, \label{e1} \feq
$$ \int_{K} d k := \int_{\reali} d \ku \int_{\reali} d \kd \sum_{\kt \in ~\{\pi/a,2 \pi/a,3 \pi/a,...\} }~; $$
\beq F_k(\bx) := {1 \over \pi \sqrt{2 a}} e^{i (\ku \xu + \kd \xd)} \sin(\kt \xt)~;
\qquad \om_k := \sqrt{\ku^2 + \kd^2 + \kt^2}~. \label{e2} \feq
The above functions fulfill $\int_{\Om} d^3 \bx \Fkc \Fh = \delta(\ku - \hu) \delta(\kd - \hd)
\delta_{\kt, \htt}$; we can use them to build $\fk(t,\bx) := \Fk(\bx) e^{- i \om_k t}$.
Let us pass to the computation of the components $\la 0 | \Tis_{\mu \nu} | 0 \ra$, and to their
analytic continuation at $\s=0$; we will start from the case $\mu=0, \nu=0$. From
Eq.s \rref{formals} and (\ref{e1}-\ref{e2}), we obtain
\parn
\vbox{
\beq \la 0 | \Tis_{0 0} | 0 \ra  \feq
$$ = {1 \over 8 \pi^2 a}
\sum_{\kt \in ~\{\pi/a,2 \pi/a,3 \pi/a,...\} }
\int_{\reali} d \ku \int_{\reali} d \kd  {\ku^2 + \kd^2 + \kt^2 - ( \ku^2 + \kd^2 + 4 \xi \kt^2)
\cos(2 \kt \xt) \over (\ku^2 + \kd^2 + \kt^2)^{\s/2 + 1/2}} $$
$$ \hspace{-1.5cm} = {1 \over 8 \pi^{\s-1} a^{4-\s}} \sum_{\ell=1}^{+\infty}
\int_{\reali} d \qu \int_{\reali} d \qd {\qu^2 + \qd^2 + \ell^2 - ( \qu^2 + \qd^2 + 4 \xi \ell^2)
\cos(2 \pi \ell \xt/a) \over (\qu^2 + \qd^2 + \ell^2)^{\s/2 + 1/2}} $$
}
where, in the last passage, we have performed a change of variables $\ku = (\pi/a) \qu$, $\kd = (\pi/a) \qd$,
$\kt= (\pi/a) \ell$. We now pass to polar coordinates in the $(\qu, \qd)$ plane, setting
$\qu = \rho \cos \theta$, $\qd = \rho \sin \theta$, and then compute the integrals in $\theta, \rho$; in
this way we obtain
$$  \la 0 | \Tis_{0 0} | 0 \ra
= {1 \over 8 \pi^{\s-1} a^{4-\s}} \sum_{\ell=1}^{+\infty}
\int_{0}^{2 \pi} d \theta \int_{0}^{+\infty} d \rho \rho {\rho^2 + \ell^2 - (\rho^2 + 4 \xi \ell^2)
\cos(2 \pi \ell \xt/a) \over (\rho^2 + \ell^2)^{\s/2 + 1/2}} $$
\beq \hspace{2.2cm} = {1 \over 4 \pi^{\s-2} (\s-3) a^{4-\s}} \sum_{\ell=1}^{+\infty}
\left[ {1 \over \ell^{\s-3}}
- {2 + 4 (\s-3) \xi \over (\s-1)} \, {\cos(2 \pi \ell \xt/a) \over \ell^{\s-3}} \right]~. \label{lastsum} \feq
The last series is clearly convergent if
\beq \Re \s > 4~; \feq
under the same condition, all the expressions given previously for $\la 0 | \Tis_{0 0} | 0 \ra$
are meaningful and finite. To go on let us recall that the
polylogarithm $(z, s)\mapsto Li_s(z)$ is defined by
\beq
Li_s(z) := \sum_{\ell=1}^{+\infty} {z^\ell \over \ell^s} \qquad
\mbox{for $z \in \complessi$, $|  z | \leqs 1$ and $s \in \Sz$}, \label{poly} \feq
where $\Sz \subset \complessi$ is the set of values of $s$ for which
the above series converges: one finds
\beq \Sz = \left\{ \barray{ll}  \complessi & \mbox{if $|z| < 1$}, \\
\{ \Re s > 0 \} & \mbox{if $|z| = 1$, $z \neq 1$}, \\ \{ \Re s > 1 \} & \mbox{if $z=1$. }
\farray \right. \label{sz} \feq
(We note that, for $|z | > 1$, there is no $s \in \complessi$ such that
the series converges.)
Let us also recall that the Riemann zeta function
$s \mapsto \zeta(s)$ is defined setting
\beq \zeta(s) := Li_s(1) = \sum_{\ell=1}^{+\infty} {1 \over \ell^s} \qquad
\mbox{for $s \in \complessi$, $\Re s > 1$}~. \label{zeta} \feq
The functions $Li$, $\zeta$ can be extended to larger domains by analytic continuation, as reviewed
in Appendix \ref{copoly}.
Comparing Eq.s \rref{lastsum} \rref{poly} \rref{zeta}, and noting that $\cos(2 \pi \ell \xt/a)
=$ $(1/2) (e^{2 i \pi \xt/a})^\ell + (1/2)  (e^{-2 i \pi \xt/a})^\ell$, we see that
\beq  \la 0 | \Tis_{0 0} | 0 \ra  \label{lasu} \feq
$$ = {1 \over 4 \pi^{\s-2} (\s-3) a^{4 - \s}}
\left\{ \zeta(\s-3) - {1 + 2 (\s-3) \xi \over (\s-1)} \,
\Big[ Li_{\s-3}(e^{2 i \pi \xt/a}) +  Li_{\s-3}(e^{-2 i \pi \xt/a}) \Big]\right\}~$$
for $\Re u > 4$. The other components $\la 0 | \Tis_{\mu  \nu} | 0 \ra$ are treated similarly. More precisely,
we find
\parn
\vbox{
\beq \la 0 | \Tis_{i i} | 0 \ra   \label{e11} \feq
$$ = {1 \over 8 \pi^2 a}
\sum_{\kt \in \{\pi/a,2 \pi/a,3 \pi/a,...\} }
\int_{\reali} d \ku \int_{\reali} d \kd  {\ki^2 - ( \ki^2 + (1 - 4 \xi) \kt^2)
\cos(2 \kt \xt) \over (\ku^2 + \kd^2 + \kt^2)^{\s/2 + 1/2}}~~(i=1,2) ; $$
\beq \la 0 | \Tis_{3 3} | 0 \ra\label{e33} \feq
$$ = {1 \over 8 \pi^2 a}
\sum_{\kt \in ~\{\pi/a,2 \pi/a,3 \pi/a,...\} }
\int_{\reali} d \ku \int_{\reali} d \kd  {\kt^2 \over (\ku^2 + \kd^2 + \kt^2)^{\s/2 + 1/2}}~; $$
\beq \la 0 | \Tis_{\mu  \nu} | 0 \ra = 0 \qquad \mbox{for $\mu \neq \nu$}~; \feq
}
indeed, one checks that $\la 0 | \Tis_{2 2} | 0 \ra =
\la 0 | \Tis_{1 1} | 0 \ra$ with a change of variables $k_2 \leftrightarrow k_1$.
The expressions \rref{e11} \rref{e33} can now be treated with the same method
employed for $\la 0 | \Tis_{0 0} | 0 \ra$: one makes a change of variables
$\ku = (\pi/a) \qu$, $\kd = (\pi/a) \qd$,
$\kt= (\pi/a) \ell$, passes to polar coordinates $(\rho, \theta)$ in the $(\qu, \qd)$ plane,
integrates in these coordinates and then expresses the remaining sum over $\ell$ in terms
of the zeta function and of the polylogarithm. The results of such computations can be summarized
in the formula
\parn
\vbox{
\beq \la 0 | \Tis_{\mu  \nu} | 0 \ra \Bigg|_{\mu,\nu = 0,1,2,3}
= A^\s
\left( \barray{cccc} \s -1 & 0 & 0 & 0 \\ 0 & 1 & 0 & 0 \\ 0 & 0 & 1 & 0 \\ 0 & 0 & 0 & \s - 3 \farray \right)
\label{formula} \feq
$$ + B^\s(\xt)
\left( \barray{cccc}
-1 - 2 (\s - 3) \xi &   0   &   0   &   0   \\
0   &   1 - \dd{\s \over 2} + 2(\s-3) \xi   &   0   &   0   \\
0   &   0   &   1 - \dd{\s \over 2} + 2(\s-3) \xi   &   0   \\
0   &   0   &   0   &   0   \farray \right) \mbox{for $\Re \s > 4$}, $$
$$ A^\s := {\zeta(\s - 3) \over 4 \pi^{\s-2} (\s - 3)(\s - 1)  a^{4 - \s}},~~
B^\s(\xt) := { Li_{\s-3}(e^{2 i \pi \xt/a}) +  Li_{\s-3}(e^{-2 i \pi \xt/a})
\over 4 \pi^{\s-2} (\s - 3)(\s - 1)  a^{4 - \s}}~. $$}
\vfill \eject \noindent
\textbf{Renormalization by analytic continuation.} Due to \rref{formula}, the problem
of the analytic continuation of $\la 0 | \Tis_{\mu  \nu} | 0 \ra$ at $\s=0$ is reduced
to the problem of continuing the functions $s \mapsto \zeta(s), Li_{s}(z)$ (for fixed $z$) up to
the point $s=-3$. As reviewed in Appendix \ref{copoly3}, such continuations are
given by
\beq Li_{-3}(z) = {z (z^2 + 4 z + 1) \over (z-1)^4}~
~~\mbox{for $z \neq 1$}; \qquad \zeta(-3) = Li_{-3}(1) = {1 \over 120}~.
\label{espl} \feq
(Note a discontinuity with respect to $z$ presented by the continuations:
$\lim_{z \vain 1} Li_{-3}(z)$ $= \infty \neq Li_{-3}(1)$. For an interpretation of this fact,
we refer again to Appendix \ref{copoly3}). \parn
Keeping in mind these facts we return to Eq. \rref{formula}, from which we infer that
$\la 0 | \Ti_{\mu  \nu} | 0 \ra_{ren} := AC \Big|_{\s=0} \la 0 | \Tis_{\mu  \nu} | 0 \ra$ is as follows:
\beq \hspace{-0.0cm} \la 0 | \Ti_{\mu  \nu} | 0 \ra_{ren} \Bigg|_{\mu,\nu = 0,1,2,3} \hspace{-0.7cm} =
A \! \left( \barray{cccc} -1 & 0 & 0 & 0 \\ 0 & 1 & 0 & 0 \\ 0 & 0 & 1 & 0 \\ 0 & 0 & 0 & - 3 \farray \right)
\! + \left(1 - 6 \xi \right) \! B (\xt) \!
\left( \barray{cccc}
-1  &   0   &   0   &   0   \\
0   &   1   &   0   &   0   \\
0   &   0   &   1   &   0   \\
0   &   0   &   0   &   0   \farray \right) \! \! . \!\!
\label{tmunuparall1} \feq
Here
$A \! := \! A^0 \! = \! \dd{\pi^2 \over 12 a^4} \! \zeta(- 3)$ and
$B(\xt) \! := \! B^0 (\xt) \!  = \! \dd{\pi^{2}
\over 12 a^4} \! \left[Li_{-3}(e^{2 i \pi \xt/a}) \!  + \!  Li_{-3}(e^{-2 i \pi \xt/a})\right]\!$,
i.e., using the expressions \rref{espl},
\beq A = {\pi^2 \over 1440 a^4}~,  \label{tmunuparall2}\feq
$$ B (\xt) = {\pi^2 \over 12 a^4}~{ 2 + \cos(2 \pi \xt/a) \over [1 - \cos(2 \pi \xt/a)]^2 } =
{\pi^2 \over 48 a^4} {3-2\sin^2 (\pi \xt/a) \over \sin^4 (\pi \xt/a)}
\qquad (0 < \xt < a)~. $$
(The first expression above for $B(\xt)$ follows using \rref{espl} with $z = e^{\pm 2 i \pi \xt/a}$;
the second expression follows from the duplication formula for the cosine). \parn
Let us remark the following: \parn
i) For $\mu=0,1,2$ the components $\la 0 | \Ti_{\mu  \mu} | 0 \ra_{ren}$ depend
on $\xt$ through the function $B$, except in the conformal case $\xi=1/6$
where they are constant. The component $\la 0 | \Ti_{3  3} | 0 \ra_{ren}$
is constant in any case. \parn
ii) The function $B(x^3)$ diverges like $1/(x^3)^4$ in the limit $x^3\to 0$,
and like $1/(x^3-a)^4$ in the limit $x^3\to a$. The same can be said
of $\la 0 | \Ti_{\mu  \mu} | 0 \ra_{ren}$ for $\mu=0,1,2$ and $\xi \neq 1/6$.
\vskip 0.1cm \noindent
Eq.s  (\ref{tmunuparall1}-\ref{tmunuparall2}) are our final result
for the renormalized stress-energy VEV. We have now checked
the following claim of the Introduction: the local zeta method,
based on analytic continuation, gives directly a finite stress-energy
tensor, \textsl{with no need to remove divergent terms}.
We already indicated this fact as a relevant
difference between this approach and the point splitting method; however
the renormalized tensors derived by these two approaches
coincide, as illustrated hereafter.
\salto
\textbf{Comparison with the results obtained by point splitting.}
Let us compare our Eq.s (\ref{tmunuparall1}-\ref{tmunuparall2}) with the results
obtained by Esposito \textsl{et al.} \cite{Esp} by the point splitting method.
The essence of this method has been reviewed in Eq.s (\ref{spli1}-\ref{spli2})
(which are implemented in \cite{Esp} using a Green function method, fully equivalent
to the eigenfunction expansion for the Laplacian). The cited work produces
the formal result
\parn
\vbox{
\beq \lim_{x' \vain x}  \la 0 | \Ti_{\mu \nu}(x,x') | 0 \ra~ =
\Bigg( A + {1 \over 2 \pi^2} \lim_{\xtp \vain \xt} {1 \over (\xt - \xtp)^4} \Bigg)
\left( \barray{cccc} -1 & 0 & 0 & 0 \\ 0 & 1 & 0 & 0 \\ 0 & 0 & 1 & 0 \\ 0 & 0 & 0 & - 3 \farray \right)
\label{tmunuesp} \feq
$$ \hspace{-0.5cm} + \left(1 - 6 \xi \right) B (\xt)
\left( \barray{cccc}
-1  &   0   &   0   &   0   \\
0   &   1   &   0   &   0   \\
0   &   0   &   1   &   0   \\
0   &   0   &   0   &   0   \farray \right) , $$
}
where $A, B$ are as in our Eq.s \rref{tmunuparall2}. As indicated in \rref{spli2},
in this apprach renormalization amounts to subtract the divergent term proportional to
$\lim_{\xtp \vain \xt} (\xt - \xtp)^{-4}$; so, the renormalized
stress-energy VEV agrees with ours. \parn
A stress-energy VEV renormalizazion, based essentially on point splitting, appears
as well in the previous book of Milton \cite{Mil} who gives for
$A$ the expression in \rref{tmunuparall2} but obtains, in place of $B$,
the function
\beq
\label{BMil}
\Bi(\xt) = {1 \over 16 \pi^2 a^4} \left[ \zeta (4, \xt /a)+\zeta (4, 1 - \xt /a )\right]~;
\feq
here $(z,s) \mapsto \zeta (s,z)$ is the Hurwitz zeta function defined by
\beq
\zeta (s,z) = \sum_{\ell = 0}^{+\infty}{1 \over (\ell + z)^s}.
\feq
Indeed, the Milton function $\Bi$ coincides with the function $B$ in Eq.
\rref{tmunuparall2}. To show this, we refer to the known identity
(see \cite{Olv}, page 608, Eq. (25.11.12))
\beq
\zeta (s+1, z) = {(-1)^{s+1} \over s!} \psi^{(s)}(z) \qquad \mbox{for $s=1,2,3,...$}~,
\feq
where the right hand
side contains the polygamma function $\psi^{(s)}(z) := ({d / d z})^{s+1} \ln \Gamma(z)$, for $s=1,2,3,...$; this implies
\beq
\Bi(\xt) = {1 \over 96 \pi^2 a^4} \left[\psi^{(3)} (\xt/a) + \psi^{(3)} (1-\xt/a) \right]~. \label{BMil2}
\feq
Another relation, known to hold for the polygamma function, is
\beq
\psi^{(s)} (1 - z) + (-1)^{s+1} \psi^{(s)} (z) = (-1)^{s} \pi {d^{s} \over dz^{s}} \cot (\pi z)
\qquad \mbox{for $s=1,2,3,...$}~; \feq
(see \cite{Olv}, page 144, Eq. (5.15.6)); this entails
\parn
\vbox{
\beq
\hspace{-1.9cm} \Bi(\xt) =
- {1 \over 96 \pi a^4} \left.\left({d^{3} \over dz^{3}} \cot (\pi z) \right)\right\vert_{z = \xt/a}
\feq
$$ \hspace{2.3cm} = {\pi^2 \over 48 a^4}\left[{3 - 2 \sin^2(\pi \xt/a) \over
\sin^4 (\pi \xt/a)} \right] = B(\xt)~\mbox{as in \rref{tmunuparall2}}~. $$
}
As a final comment on this result, we mention that the equality
$\Bi = B$ is a special case of a more general
relation between the polylogarithm and the Hurwitz zeta function
(see \cite{Jon}, or \cite{Olv} for a reformulation in modern
notations).
\section{The Casimir effect outside one plate, or two parallel plates.}
\label{caout}
\textbf{The case of a single plate.}
Let the plate
occupy the plane $\xt = 0$; heferafter we determine the renormalized VEV of $\Ti_{\mu \nu}$ in
one of the half-spaces bounded by the plane, say, in
\beq \Omp := \{ (\xu, \xd, \xt)~|~\xu, \xd \in \reali~, \xt > 0 \}~.\feq
As before, we assume for the (scalar, gravity coupled) field $\Fi$ the Dirichlet boundary conditions
\beq \Fi(t,\xu, \xd, \xt) = 0 \qquad \mbox{for $\xt=0$}~. \feq
To treat this case, it is not even necessary to set up a framework as in the previous sections,
starting from the Dirichlet eigenfunctions of $\Delta$ in $\Omp$. In fact,
it suffices to view $\Omp$ as the $a \vain + \infty$ limit of the domain
\beq \Oma := \{  (\xu, \xd, \xt)~|~\xu, \xd \in \reali~, 0 < \xt <  a \} \feq
and define the renormalized VEV of $\Ti_{\mu \nu}$ in $\Omp$ as
\beq \la 0 | \Ti_{\mu  \nu} | 0 \ra_{\infty, ren} := \lim_{a \vain + \infty}
\la 0 | \Ti_{\mu  \nu} | 0 \ra_{a, ren}~, \feq
where the right hand side contains the renormalized VEV in $\Oma$; the latter is known
from the previous section, see Eq.s (\ref{tmunuparall1}-\ref{tmunuparall2}). So,
\parn
\vbox{
\beq \la 0 | \Ti_{\mu  \nu} | 0 \ra_{\infty, ren} = \Bigg( \lim_{a \vain + \infty} A_a \Bigg)
\left( \barray{cccc} -1 & 0 & 0 & 0 \\ 0 & 1 & 0 & 0 \\ 0 & 0 & 1 & 0 \\ 0 & 0 & 0 & - 3 \farray \right)  \feq
$$ + \left(1 - 6 \xi \right) \! \Bigg( \lim_{a \vain + \infty} B_a(\xt) \Bigg) \!\!
\left( \barray{cccc}
-1  &   0   &   0   &   0   \\
0   &   1   &   0   &   0   \\
0   &   0   &   1   &   0   \\
0   &   0   &   0   &   0   \farray \right)\! ,~~
\mbox{$A_a := A$, $B_a(\xt) := B(\xt)$ as in \rref{tmunuparall2}}. $$
}
\noindent
From Eq. \rref{tmunuparall2}, it is evident that
$A_a \vain 0$, $B_{a} (\xt) \vain 1/(16 \pi^2 (\xt)^4)$ for $a \vain + \infty$; so,
\beq \la 0 | \Ti_{\mu  \nu} | 0 \ra_{\infty, ren} =
{1 - 6 \xi \over 16 \pi^2 (\xt)^4}
\left( \barray{cccc}
-1  &   0   &   0   &   0   \\
0   &   1   &   0   &   0   \\
0   &   0   &   1   &   0   \\
0   &   0   &   0   &   0   \farray \right) \qquad (0 < \xt < + \infty)~. \label{fires} \feq
The above result, derived in a different way, appears e.g.
in \cite{Saha}. Of course, one obtains similar conclusions in the half space
$\{ -\infty < \xt < 0 \}$. \parn
We observe that, if the coupling parameter takes the conformal value $\xi = 1/6$,
the stress-energy tensor vanishes everywhere outside the plate. \salto
\textbf{The case outside two parallel plates.}
We now consider, as in the previous section, two plates
occupying the planes $\xt=0$ and $\xt = a$; we are interested
in the renormalized VEV of $\Ti_{\mu \nu}$ in the region
outside the plates, which is the disjoint union of the half
spaces $\{ \xt < 0 \}$ and $\{ \xt > a \}$. This can be
obtained by obvious adaptations of the result \rref{fires} on the half
space $\{ \xt > 0 \}$; the conclusion is
\beq \la 0 | \Ti_{\mu  \nu} | 0 \ra_{ren} =
{1 - 6 \xi \over 16 \pi^2 (\xt)^4}
\left( \barray{cccc}
-1  &   0   &   0   &   0   \\
0   &   1   &   0   &   0   \\
0   &   0   &   1   &   0   \\
0   &   0   &   0   &   0   \farray \right) \qquad (-\infty < \xt < 0)~. \label{fires1} \feq
\beq \la 0 | \Ti_{\mu  \nu} | 0 \ra_{ren} =
{1 - 6 \xi \over 16 \pi^2 (\xt - a)^4}
\left( \barray{cccc}
-1  &   0   &   0   &   0   \\
0   &   1   &   0   &   0   \\
0   &   0   &   1   &   0   \\
0   &   0   &   0   &   0   \farray \right) \qquad (a < \xt < +\infty)~. \label{fires2} \feq
Note that, in the conformal case $\xi=1/6$,
$\la 0 | \Ti_{\mu  \nu} | 0 \ra_{ren}$ is identically zero outside the plates.
If $\xi \neq 1/6$, the components with $\mu=\nu=0,1,2$ of this tensor
diverge like $1/(\xt)^4$ and $1/(\xt-a)^4$ for $\xt \vain 0^{-}$
and $\xt \vain a^{+}$, respectively; we recall that
similar divergences were found as well for the stress-energy tensor
between the plates. \parn
\salto
\section{Pressure on the plates}
\label{press}
In this section we alway use the spatial indices $i,j \in \{1,2,3\}$.
Let us consider any one of the two plates at $\xt=0$ or $\xt = a$,
and evaluate the force per unit area acting on it;
in principle, this computation should take
into account the action of the field both inside and
outside the plates.
The force per unit area produced on the given plate
by the field in the inner region is $p^i_{in} = \la 0 | \Ti^{i}_{~j} | 0 \ra_{in} n^j_{out}$
where $\la 0 | \Ti^{i}_{~j} | 0 \ra_{in}$ is the renormalized
stress-energy tensor in the inner region and $n^j_{out}$ the normal
unit vector to the plate pointing towards the outer region.
On the other hand, the force per unit area produced on the same
plate by the field in the outer region is
$p^i_{out} = \la 0 | \Ti^{i}_{~j} | 0 \ra_{out} n^j_{in}$, where
the subscripts $_{in}$, $_{out}$ have an obvious meaning. So, the total
force per unit area on the plate is
\beq p^i = \la 0 | \Ti^{i}_{~j} | 0 \ra_{in} n^j_{out} + \la 0 | \Ti^{i}_{~j} | 0 \ra_{out} n^j_{in}~. \feq
For the plate located at $\xt = 0$, we have $(n^j_{in}) = (0,0,1)$, $(n^j_{out}) = (0,0,-1)$,
so
\beq p^i \Big|_{\xt=0} =  \la 0 | \Ti^{i}_{~3} | 0 \ra_{out}  - \la 0 | \Ti^{i}_{~3} | 0 \ra_{in}~\Big|_{\xt=0}; \feq
for the plate at $\xt = a$ the inner ond outer normals are reverted, so
\beq p^i \Big|_{\xt=a} =  - \la 0 | \Ti^{i}_{~3} | 0 \ra_{out}  + \la 0 | \Ti^{i}_{~3} | 0 \ra_{in}~\Big|_{\xt=a}~. \feq
Now, we take the expressions of
$\la 0 | \Ti^{i}_{~3} | 0 \ra_{in, out} = \la 0 | \Ti_{i 3} | 0 \ra_{in, out}$
from Eq.s (\ref{tmunuparall1}-\ref{tmunuparall2}) and (\ref{fires1}-\ref{fires2});
for both plates  $\la 0 | \Ti_{i 3} | 0 \ra_{out}$ vanishes
and $(\la 0 | \Ti_{i 3} | 0 \ra_{in}) = (0,0, - 3 A)$ with $A = {\pi^2/1440 a^4}$,
as usually; in conclusion
\beq \left( p^i \Big|_{\xt=0} \right) = (0,0, {\pi^2 \over 480 a^4})~; \feq
\beq \left( p^i \Big|_{\xt=a} \right) = (0,0, - {\pi^2 \over 480 a^4})~. \feq
Thus the plates are subject to a reciprocal attraction inverserly
proportional to the fourth power of their distance.
We note that, once more, the result obtained
agrees with the ones reported in \cite{Esp, Mil}.
\vskip 0.7cm \noindent
\textbf{Acknowledgments.} This work was partly supported by INdAM, INFN and by MIUR, PRIN 2008
Research Project ``Geometrical methods in the theory of nonlinear waves and applications". \parn
We gratefully acknowledge Giuseppe Molteni for useful indications about the polylogarithm.
\vfill \eject \noindent
{~}
\vskip -2cm
\noindent
\appendix
\section{Appendix. Analytic continuation of the polylogarithm (and
of the zeta function).}
\label{copoly}
Let us report Eq.s \rref{poly} \rref{zeta}
$$
Li_s(z) := \sum_{\ell=1}^{+\infty} {z^\ell \over \ell^s}~
\mbox{for $z \in \complessi$, $|z| \leqs 1$ and $s \in S_z$}; $$
$$ \zeta(s) := Li_s(1) := \sum_{\ell=1}^{+\infty} {1 \over \ell^s}~
\mbox{for $s \in \complessi$, $\Re s > 1$}~. $$
(In the above $\Sz$ is the subset of $\complessi$ such that the series
for $Li_s(z)$ converges, see Eq. \rref{sz}).
Our problem is continuing analytically (in $s$) the functions defined as above;
the solution is well known \cite{Jon}, and reported here for completeness. Indeed, let us
define
\beq Li_s(z) := - {\Gamma(1 - s)  \, z \over 2 \pi i} \int_{H_z} d t {(-t)^{s-1} \over e^{t}- z}
\qquad \mbox{for $z \in \complessi$,
$s \in \complessi \setminus \{1,2,3,..\}$}~, \label{polye} \feq
\beq {~} \hspace{-0.5cm} Li_s(z) := \lim_{s' \vain s} Li_{s'}(z)
~\mbox{for $z \in \complessi \setminus \{1\}$, $s \in \{1,2,3,...\}$
or $z=1$, $s \in \{2,3,...\}$}; \label{polyen} \feq
\beq \zeta(s) := Li_s(1) \qquad \mbox{for $s \in \complessi \setminus \{1 \}$}~. \label{zetae} \feq
In Eq. \rref{polye}, $\Gamma$ is the usual Gamma function. Furthermore: \parn
i) $H_z$ is a Hankel contour in the complex $t$ plane,
starting from infinity in the direction of
the positive real axis, turning counterclockwise around $t=0$
and returning to infinity in the direction of the
positive real axix (see the figure below);
this contour is chosen so that \textsl{all} the solutions $t$ of the equation $e^t = z$ are
outside the region bounded by $H_z$, except the solution $t=0$ appearing if $z=1$. \parn
ii) For each $t \in H_z$ we intend
\beq (-t)^{s-1} := e^{-i (s-1) \pi} t^{s-1}~, \qquad t^{s-1} := |t|^{s-1} e^{i (s-1) \arg t} \feq
where $t \mapsto \arg t$ is the unique continuous function on $H_z$
such that $\arg t \vain 0$ when $t$ tends to the beginning of the path. \parn
\vskip -0.5cm \noindent
\begin{figure}[h] \centering
\includegraphics[width=9cm]{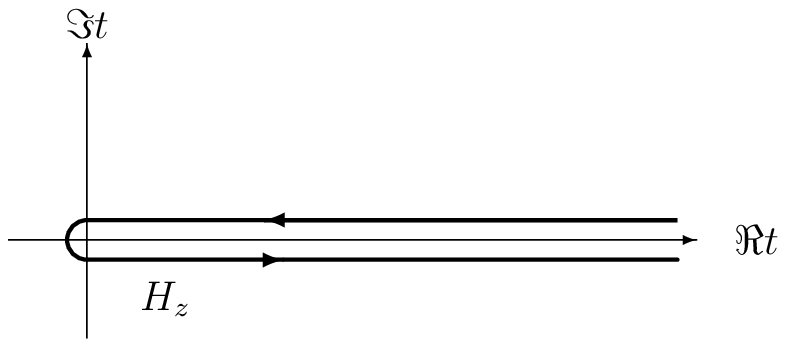}
\end{figure}
\vfill \eject \noindent
For each fixed $z$ with $|z| \leqs 1$,
the function $s \mapsto Li_s(z)$ defined via
\rref{polye} \rref{polyen} is
the (unique) analytic continuation of the function $s \in \Sz \mapsto Li_s(z)$
previously defined via the power series \rref{poly}; to prove this,
it suffices to prove that the definition \rref{polye} for $Li_s(z)$ via
a contour integral implies
a series expansion as in \rref{poly}, if $s \in \Sz$. To this purpose,
we reexpress the function of $z$ and $t$ in \rref{polye} in the following way:
\beq z {(-t)^{s-1} \over e^t - z} = (-t)^{s-1} {z e^{-t} \over 1 - z e^{-t}} =
(-t)^{s-1} \sum_{\ell=1}^{+\infty} (z e^{-t})^\ell~; \label{seresp} \feq
inserting this result into Eq. \rref{polye}, we obtain
\beq Li_s(z) = - {\Gamma(1-s) \over 2 \pi i} \sum_{\ell=1}^{+\infty} z^\ell
\int_{H_z} (-t)^{s-1} e^{-\ell \, t}~. \label{previli} \feq
On the other hand, the known Hankel's integral representation
for $1/\Gamma$ \cite{Olv} implies
\beq - {1 \over 2 \pi i} \int_{H_z} (-t)^{s-1} e^{-\ell \, t} = {1 \over \Gamma(1-s) \ell^s} \feq
for $\ell=1,2,3...~$.
The last two equations yield the wanted expansion
$Li_s(z) = \sum_{\ell=1}^{+\infty} z^\ell/\ell^s$, of the form \rref{poly}.
\parn
The above manipulations have hidden a problem: to grant convergence
of the series expansion \rref{seresp} and the exchange
between the summation over $\ell$ and the integration over $H_z$,
one shoud have $|z e^{-t}| < 1$ uniformly in $t \in H_z$:
on the other hand, $|z e^{-t}|$ can be larger than $1$ when $t$
is on the arc in the half plane $\Re t < 0$, turning around the origin.
Let us skecth how to overcome this difficulty; the basic idea
is that $Li_s(z)$ defined in \rref{polye} does not change if
we shrink the path $H_z$ around the origin. If $|z| < 1$,
we can shrink $H_z$ so that $|z e^{-t}| < 1$ uniformly
in $H_z$, including the arc that turns around the origin.
The case $|z| = 1$ is a bit more technical: one
isolates from the integral over $H_z$ the contribution
of the arc encircling the origin, makes a series
expansion of the integrand in the remaining part of $H_z$, and
finally proves that the contribution from the arc
can be made arbitrarily small by shrinking.
\parn
The previous results on the analytic continuation
of the function $s \mapsto Li_s(z)$ hold, in particular,
for $z=1$; so, the function $s \in \complessi \setminus \{1 \}
\mapsto \zeta(s)$ in \rref{zetae} is the (unique)
analytic continuation of the function defined previously by \rref{zeta}. \parn
Another property of the function \rref{polye}\rref{polyen} is that it is jointly analytic
in $(z, s)$, when these variables range in a suitable open subset of $\complessi^2$;
outside this open set, some pathologies can appear. In particular, for a given $s$,
this function can happen to be discontinuous in $z$
at the specific point $z=1$: see, e.g., the case $s=-3$
discussed hereafter. \parn
\section{Appendix. The polylogarithm (and the zeta function)
at $\boma{s=-3}$.}
\label{copoly3}
Let us consider the analytic continuation of the polylogarithm
described in Appendix \ref{copoly}, and evaluate it at $s=-3$.
For this choice of $s$,
Eq.s \rref{polye} takes the form
\beq Li_{-3}(z) = - {6 \, z \over 2 \pi i} \int_{H_z} {d t \over t^4 (e^{t}- z)}
\qquad \mbox{for $z \in \complessi$}~; \label{polye3} \feq
the integral therein is easily
computed by the method of residues, as briefly skecthed hereafter.
First of all, the integral in \rref{polye} involves a meromorphic function
of $t$, whose only singularity in the region
bounded by $H_z$ is a pole at $t=0$. The order of the pole is
$4$ if $z\neq 1$, while it is $5$ if $z=1$, and one finds
$$ \mbox{Res} \left[{1\over t^4 (e^t - z)} \right]_{t=0} =
- {(z^2 + 4 z + 1) \over 6 (z-1)^4}~~\mbox{if $z \neq 1$};~~
\mbox{Res} \left[{1\over t^4(e^t - 1)} \right]_{t=0} = - {1 \over 720}~. $$
When these results are inserted into \rref{polye3},
the residue theorem gives
$$ Li_{-3}(z) = {z (z^2 + 4 z + 1) \over (z-1)^4}~
\mbox{for $z \in \complessi \setminus \{1\}$};~~\zeta(-3) = Li_{-3}(1) = {1 \over 120}~; $$
these are the statements \rref{espl}, which are now justified. \parn
To conclude, we note a discontinuity of the type mentioned at the end
of Appendix \ref{copoly}:
$\lim_{z \vain 1} Li_{-3}(z) = \infty \neq Li_{-3}(1)$.
This is basically due to the jump in the order of the pole
(from $4$ to $5$) when $z$ goes to $1$.
\vfill \eject 
\end{document}